\documentclass{aa}  
\usepackage{graphicx}
\usepackage{txfonts}
\newcommand{\lsim}{\raisebox{-0.13cm}{~\shortstack{$<$ \\[-0.07cm] $\sim$}}~}
\newcommand{\gsim}{\raisebox{-0.13cm}{~\shortstack{$>$ \\[-0.07cm] $\sim$}}~}
\newcommand{\tfm}{$\rm 24 \, \mu m \,$}

\begin{document}
   \title{The role of the LIRG and ULIRG phases in the evolution of $K_s$-selected galaxies}

   \subtitle{}

   \author{K. I. Caputi\inst{1},  H. Dole\inst{1}, G. Lagache\inst{1},
            R. J. McLure\inst{2}, J. S. Dunlop\inst{2}, J.-L. Puget\inst{1},  
	    E. Le Floc'h\inst{3,4}, and P. G. P\'erez-Gonz\'alez\inst{3}}

   \offprints{K.I. Caputi; e-mail: kcaputi@ias.u-psud.fr}

   \institute{Institut d'Astrophysique Spatiale (IAS), b\^atiment 121, F-91405 Orsay, France; Universit\'e Paris-Sud 11 and CNRS (UMR 8617)\\
              \email{kcaputi,herve.dole,guilaine.lagache,puget@ias.u-psud.fr}
         \and
             Institute for Astronomy, University of Edinburgh, Royal Observatory, Blackford Hill, EH9 3HJ,                    Edinburgh, UK\\
             \email{rjm,jsd@roe.ac.uk}
	  \and
	      Steward Observatory, University of Arizona, 933 N Cherry Ave, Tucson, AZ 85721, USA\\
	      \email{elefloch,pgperez@as.arizona.edu}   
	   \and
	      Also associated to the Observatoire de Paris, GEPI, 92195, Meudon, France   
             }

   \date{Received ...; accepted ...}

 
  \abstract
   {}
   {We investigate the role of the luminous infrared galaxy (LIRG) and ultra-luminous infrared galaxy (ULIRG)  phases in the evolution of $K_s$-selected galaxies and, in particular, Extremely Red Galaxies (ERGs).}
   {With this aim, we compare the properties of a sample of 2905 $K_s<21.5$ (Vega mag) galaxies in the GOODS/CDFS with the sub-sample of those 696 sources which are detected at \tfm.}
   {We find that LIRGs constitute 30\%  of the galaxies with stellar mass $M>1\times10^{11} \, M_{\odot}$ assembled at redshift $z=0.5$. A minimum of 65\% of the galaxies with  $M>2.5\times10^{11} \, M_{\odot}$ at $z \approx 2-3$ are ULIRGs at those redshifts.  60\% of the ULIRGs in our sample have the characteristic colours of ERGs. Conversely, 
 40\% of the ERGs with stellar mass  $M>1.3 \times10^{11} \, M_{\odot}$ at $1.5<z<2.0$ and a minimum of 52\% of those with the same mass cut at $2.0<z<3.0$ are ULIRGs. The average optical/near-IR properties of the massive ERGs at similar redshifts that are identified with ULIRGs and that are not have basically no difference, suggesting that both populations contain the same kind of objects in different phases of their lives. }
   {LIRGs and ULIRGs have an important role in galaxy evolution and mass assembly,
  and, although they are only able to trace a fraction of the massive ($M>1\times10^{11} \, M_{\odot}$) galaxies present in the Universe at a given time, this fraction becomes very significant ($\gsim 50$\%) at redshifts $z \gsim 2$.}

   \authorrunning{Caputi et al.}
   \titlerunning{The role of LIRGs and ULIRGs}
   \maketitle
%

\section{Introduction}

   The study of the evolution of near-infrared (IR)-selected galaxies 
   is probably the most efficient way to trace the build up of stellar 
   mass with redshift (e.g. Dickinson et al.~\cite{dickinson03};
   Fontana et al.~\cite{fontana04}; 
   Glazebrook et al.~\cite{glazeb04}; 
   Caputi et al.~\cite{cap05}, 2006a (\cite{cap06a})). In contrast to other
   wavelength surveys, near-IR observations are able to detect
   systems  with different dust content and star-formation histories.
   Thus, they appear as a viable method to produce a relatively  unbiased
   census of galaxy populations, from which the history of stellar mass 
   assembly can be investigated.      
   
   The {\em Spitzer Space Telescope} (Werner et al.~\cite{werner04})
   is making possible the 
   mapping of the  mid-IR Universe from low to high redshifts. 
   Recent studies have exploited
   this capability to set tight constraints on the intensity and the 
   composition of the mid and far-IR backgrounds (Papovich et al.~ \cite{pap04};
   Le Floc'h et al.~\cite{lefl05}; P\'erez-Gonz\'alez et al.~\cite{pg05};
   Caputi et al.~2006b (\cite{cap06b}); Dole et al.~\cite{dole06}).
   The analysis of the evolution of mid-IR
   galaxies with redshift provides fundamental information to understand  
   how star-formation and quasar activity proceeded through cosmic time.

   It is so far not  clear, however, what the role of  bright mid-IR-selected
   galaxies is  within the history of stellar mass assembly.
   The aim of this work is to assess the significance of this role, through the study  
   of mid-IR-selected galaxies in the context of the evolution 
   of the $K_s$-band galaxy population. With this purpose, we focus on the analysis of
   two main topics: 1) the number densities of massive luminous mid-IR-selected
   galaxies, in comparison to the total number densities of massive galaxies in the $K_s$-band
   population,  at different redshifts, and 2) the relation between the luminous IR  
   and the Extremely Red Galaxy (ERG) phases.  We adopt throughout 
   a cosmology with $\rm H_o=70 \, km \, s^{-1} \, Mpc^{-1}$, $\rm \Omega_M=0.3$ 
   and $\rm \Omega_\Lambda=0.7$. A Salpeter IMF over stellar masses $M=(0.1-100) \, M_\odot$ is assumed.


\section{The samples}
   
\begin{figure}
\centering
\includegraphics[width=8cm]{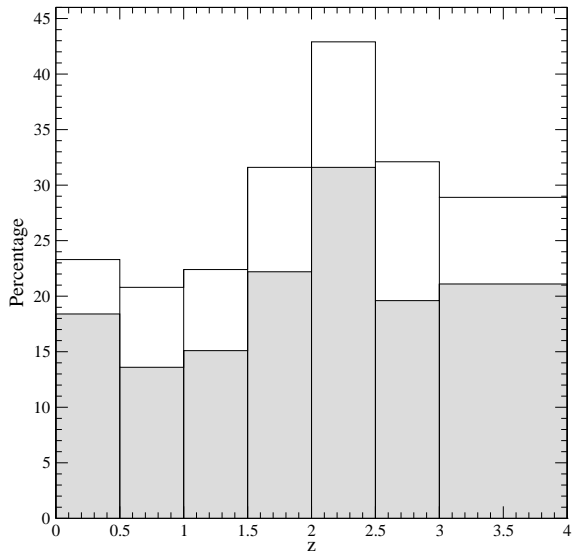}
\caption{The percentage of  $K_s<21.5$ mag galaxies which are detected in the deep MIPS/GTO \tfm catalogue, as a function of redshift. The shaded and empty histograms indicate the percentages of galaxies with $S_\nu(24 \, \rm \mu m)>83 \, \rm \mu Jy$ and the total of \tfm-detected galaxies, respectively. The faintest galaxies considered here have $S_\nu(24 \, \rm \mu m) \sim 20 \, \rm \mu Jy$, but galaxies with $S_\nu(24 \, \rm \mu m) < 60 \, (83) \, \rm \mu Jy$ only constitute $\sim 11 \, (30)\%$ of the total sample.}
\label{figperc}
\end{figure}

   ~\cite{cap06a} selected a sample of 2905 $K_s<21.5$ (Vega mag) galaxies in 131 arcmin$^2$ of the 
   Great Observatories Origins Deep Survey (GOODS) Chandra Deep Field South
   (CDFS). This area benefits from deep $J$ and $K_s$-band coverage by the 
   Infrared Spectrometer and Array Camera (ISAAC) on the Very Large Telescope
   (VLT). Other deep multiwavelength data also exist in this field; in particular, 
   $B, V , i$ and $z$-band imaging from the Advanced Camera for Surveys (ACS)
   on-board the Hubble Space Telescope (HST), and $3.6$ to $8.0 \, \rm \mu m$
   imaging from the Infrared Array Camera (IRAC) on board {\em Spitzer}.

   ~\cite{cap06a} constructed an optimized redshift catalogue for the total $K_s<21.5$ mag sample,
   incorporating the available spectroscopic (e.g. Le F\`evre et al.~\cite{lefev04}; Vanzella et al.~\cite{van05})  and COMBO17 (Wolf et al.~\cite{wolf04}) redshifts    for the CDFS.  The remaining redshifts were obtained with the public codes HYPERZ (Bolzonella, Miralles \& Pell\'o~\cite{bolz00}) and BPZ (Ben\'{\i}tez~\cite{ben00}), using multiwavelength photometry.  The comparison of the redshifts for the sources with spectroscopic data showed that the photometric     estimates had very good quality, with a median of 
   $|z_{spec}-z_{phot}|/ (1+z_{spec})=0.02$. The code HYPERZ has simultaneously been  used to model the spectral energy distribution (SED) of each source, with the Calzetti et al.~(\cite{calz00}) reddening law to take into account dust-corrections.

   The $K_s<21.5$ mag sample is cleaned of galactic stars. It also excludes those X-ray sources whose SEDs could not be satisfactorily fitted with the HYPERZ stellar templates,  as their optical/near-IR light could be contaminated by an active (AGN/QSO) component. For these sources, no  reliable mass estimate can be obtained from the modelled SED. The rejected X-ray sources constitute $\sim 2$\%  of the total $K_s<21.5$ mag sample. We refer the reader to \cite{cap06a} for further details on the selection and analysis of the $K_s<21.5$ mag galaxy sample.

   Observations of the CDFS have also been carried out with the Multiband Imaging Photometer
   for {\em Spitzer} (MIPS; Rieke et al.~\cite{rieke04}), as part of the Guaranteed Time Observers      (GTO) program. The MIPS GTO \tfm catalogue achieves 80\% completeness at a flux
   limit of $S_\nu(\rm 24 \, \mu m) \approx 83 \, \rm \mu Jy$ (see details in Papovich et al.~\cite{pap04}).

   ~\cite{cap06b} cross-correlated the \cite{cap06a} sample (and all the excluded $K_s<21.5$ mag stars  and AGN/QSO) with the deep \tfm catalogue for the CDFS.  $\sim 94\%$ of the \tfm objects with  flux $S_\nu(\rm 24 \, \mu m) > 83 \, \rm \mu Jy$ were identified with a $K_s<21.5$ mag  counterpart.   This allowed for the characterization of nearly all the brightest sources composing the \tfm background in the GOODS/CDFS.

   The fraction of \tfm galaxies with multiple $K_s$-band associations is small. ~\cite{cap06b} found that, within a matching radius of 2 arcsec, $<8\% \,(0.3\%)$ of the \tfm galaxies could be associated with  two (three) different $K_s$-band galaxies. On the other hand, $\sim 95\%$ of the associations can be done restricting the matching radius to 1.5 arcsec. Using this smaller matching radius, the number of double identifications is only $<3\%$. In all cases, ~\cite{cap06b} considered that the closest $K_s$-band source to each \tfm galaxy was the real counterpart.

   In addition, using empirical relations between the mid-IR and bolometric IR luminosities (Chary \& Elbaz~\cite{chel01}, Elbaz et al.~\cite{elbaz02}), \cite{cap06b} obtained IR luminosity ($L_{IR}$) estimates and derived star-formation rates ($SFR$) for most of their galaxies. 
In agreement with other works (Le Floc'h et al.~\cite{lefl05}; P\'erez-Gonz\'alez et al.~\cite{pg05}), they found that the mid-IR output was dominated by luminous IR galaxies (LIRGs, $10^{11} \, L_\odot < L < 10^{12} \, L_\odot$) at redshift $z \sim 1$ and by ultra-luminous IR galaxies (ULIRGs, $L > 10^{12} \, L_\odot$) at redshift $z \gsim 1.5-2.0$.
    
   In this work we analyze the role of the sub-sample of the $K_s<21.5$ mag galaxies identified with  a \tfm counterpart, within the total $K_s<21.5$ mag galaxy population.  We quantify the importance of galaxies with different IR luminosities  in the evolution of massive systems at different redshifts. The sub-sample of  \tfm-detected galaxies contains 696 out of 2905 $K_s<21.5$ mag galaxies. 
   
   The percentage of the $K_s<21.5$ mag galaxies which are \tfm-detected as a function of redshift is shown in Figure~\ref{figperc}. The shaded and empty histograms indicate the percentages of galaxies with $S_\nu(24 \, \rm \mu m)>83 \, \rm \mu Jy$ and the total of the \tfm-detected galaxies, respectively. The faintest sources included in our catalogue have fluxes $S_\nu(24 \, \rm \mu m) \sim 20 \, \rm \mu Jy$,  but galaxies with $S_\nu(24 \, \rm \mu m) < 60 \, (83) \, \rm \mu Jy$ only constitute $\sim 11 \, (30)\%$ of the total sample.  The number of spurious sources in the \tfm catalogue becomes non-negligible at fluxes $S_\nu(24 \, \rm \mu m)<83 \, \rm \mu Jy$ and starts to be important  at $S_\nu(24 \, \rm \mu m) \lsim 60 \, \rm \mu Jy$ (Papovich et al. 2004). Nevertheless, we note that the reliability of all the \tfm sources analyzed here is assured  by our $K_s$ and other band detections. Even so, there still could be the possibility of a false \tfm source which is associated just by chance with a $K_s$ counterpart.  However, we mentioned above that the fraction of \tfm sources with a double association within 1.5 arcsec is $<3\%$. We can consider this percentage to be representative of the maximum fraction of false identifications. Thus, we estimate that the fraction of spurious $S_\nu(24 \, \rm \mu m)<60 \, \rm \mu Jy$ sources associated just by chance with a $K_s$ galaxy must constitute $< 0.03 \times 11\%$=0.33\% of our total sample. Then, those plausible spurious sources do not constitute a concern in this work.

 We see in Figure~\ref{figperc} that $\sim 20-25$\% of our $K_s<21.5$ mag galaxies at redshifts $z < 1.5$ are \tfm-detected at the depth of the MIPS/GTO CDFS catalogue. The percentage of \tfm-detected galaxies increases at higher redshifts, with a maximum of $\sim 43$\% at $z \sim 2.0-2.5$. The magnitude limit of our $K_s$-band survey imposes that we only see massive galaxies at high redshifts. Above redshifts $z=2$  and $z=3$, we are only strictly complete for stellar masses $M>6\times10^{10} \, M_{\odot}$  and $M>1.3\times10^{11} \, M_{\odot}$, respectively (cf. \cite{cap06a} and Figure \ref{masslumcompl}). Thus, the  increasing percentage of IR  galaxies with redshift in particular implies that the fraction of massive galaxies with IR-activity has been significantly larger in the past. We further discuss this issue in Section \ref{secrole}. However, it should also be noted that this positive selection effect on high-redshift galaxies is very probably favoured  by the presence of policyclic aromatic hydrocarbon (PAH) emission features entering the  \tfm-filter passband (cf.  \cite{cap06b}). PAH emission lines characterize the interstellar medium of star-forming regions (D\'esert, Boulanger \& Puget~\cite{des90}). The presence of PAHs implies  that a substantial fraction of the $K_s<21.5$ mag galaxies at $z \sim 2.0-2.5$ have IR activity mainly due to star-formation. PAH emission in high-z galaxies has been spectroscopically confirmed in different recent works (e.g. Houck et al.~\cite{ho05}; Yan et al.~\cite{yan05}; Lutz et al.~\cite{lutz05}).


\section{The normal- versus active-galaxy separation}
\label{secsep}

\begin{figure}
\centering
\includegraphics[width=8cm]{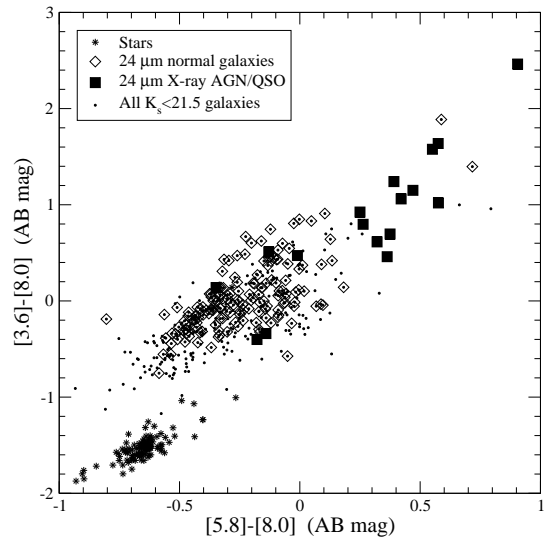}
\caption{IRAC-based colour-colour diagram for the $K_s<21.5$ mag sources with redshifts $z>1.5$. In this plot, we label as `normal' to all those galaxies which are not X-ray classified as AGN/QSO. Galactic stars have also been added for a comparison.}
\label{figiraccol}
\end{figure}

    As we mentioned above, those X-ray sources with no satisfactory modelled SED have been left out of the $K_s<21.5$ mag sample. This implies that 40 X-ray-detected \tfm sources (including X-ray AGN/QSO and other galaxies with undetermined X-ray classification) are excluded from the sub-sample analyzed here. However, we note that our \tfm sub-sample still contains  26 sources X-ray classified as AGN/QSO (with satisfactory HYPERZ SED fits).  
    
  To test the completeness of the X-ray detections to separate all the \tfm-detected active galaxies, we construct a colour-colour diagram based on different IRAC passbands for the high redshift ($z>1.5$) sources. The separation of normal and active galaxies in such a diagram should be produced by the mid-IR excess characterising the latter type of objects. 
  At lower redshifts, an AGN/QSO separation based only on the X-ray classification is expected to be more complete than at higher redshifts. On the other hand, the existence of star-forming galaxies with PAH emission lines within the IRAC passbands complicates the colour-colour separation at low redshifts, as these emission lines might mimic the AGN IR colour excesses. Thus, we restrict the analysis of the colour-colour diagram to  sources with $z>1.5$. Figure \ref{figiraccol} shows the location in this diagram of different types of \tfm sources and all the $K_s<21.5$ mag sources, at  $z>1.5$. We clearly see from this diagram that the majority ($\sim 70\%$) of the \tfm-detected X-ray AGN/QSO lie in a segregated region, typically with colours $[5.8]-[8.0 \, \rm \mu m] \gsim 0.2$ (AB mag). A few apparently normal (i.e. non X-ray classified AGN/QSO) galaxies appear within this region. The vast majority of normal galaxies occupy a different locus in the colour-colour diagram. However, we note that the remaining $\sim 30\%$ of the \tfm-detected X-ray-classified AGN/QSO also have the colours characterising  normal galaxies. Thus, we conclude that  the application of a $[5.8]-[8.0 \, \rm \mu m]$ colour cut can be useful to complement  the X-ray  AGN classification,  but cannot be both complete and reliable at the same time (cf. e.g. Barmby et al. 2006).
  
  In order to  achieve a balance between completeness and reliability  for the active galaxy identifications within our sample, we adopt the following criteria for our \tfm-detected $K_s<21.5$ mag sources. At redshifts $z<1.5$, we base our AGN/QSO classification exclusively on the X-ray data. For sources at higher ($z>1.5$) redshifts, we considered as AGN/QSO to all those objects classified as such in the X-rays, plus the two additional \tfm sources which also have a colour $[5.8]-[8.0 \, \rm \mu m] > 0.2$ (AB mag; cf. figure \ref{figiraccol}). In addition, we also classify as active galaxies eight additional objects of our sample which have been identified as AGN in the catalogue constructed by Alonso-Herrero et al.~(\cite{alon06}; table 1 of this paper). Thus, using all these criteria, the total number of sources classified as AGN/QSO in our \tfm-detected $K_s<21.5$ mag sample is $26+2+8=36$.  Throughout this paper, when we refer to the AGN/QSO of our sample, we always consider all of these 36 objects, unless otherwise explicitly stated.

\begin{table*}
\caption{The comoving number densities of massive galaxies with bolometric IR luminosity  $L_{IR}>10^{11}\, L_\odot$ (LIRGs and ULIRGs) versus redshift. The redshift bins indicated with an asterisk are not complete for this luminosity cut, so the corresponding number densities should be taken as lower limits. The percentages given in the columns \%($z$) refer to the fraction of objects classified as LIRGs or ULIRGs among all the $K_s<21.5$ mag galaxies in that redshift bin and with the same mass cut. The percentages \%($z=0$) are obtained by dividing the number density of LIRGs and ULIRGs in that redshift/mass bin by the number density of galaxies with the same mass cut
in the local Universe. These numbers indicate what fraction of the massive galaxies we see today have been IR-active in the past.  Between brackets, we show the corrected values obtained when excluding the known AGN/QSO. These AGN/QSO are only those remaining in the \tfm-detected $K_s<21.5$ mag sub-sample considered here (i.e. with satisfactory HYPERZ SED fits and, thus, reliable mass estimates), which have been classified following the criteria explained in Section~\ref{secsep}. Therefore, they constitute a lower limit to the total number of mid-IR detected active galaxies. 
} 
\label{tcnd11}      
\centering          
\begin{tabular}{c r c r c r c r c}      
\hline\hline
 & \multicolumn{4}{c}{$\rm M>1.0\times10^{11} \,M_\odot$} & \multicolumn{4}{c}{$\rm M>2.5\times10^{11} \,M_\odot$} \\
\hline 
$z$  & $\rm \rho_c (\times 10^{-5}\, Mpc^{-3})$ & \%($z$) & \% AGN($z$)&\%($z=0$) & $\rm \rho_c (\times 10^{-5}\, Mpc^{-3})$ & \%($z$) & \% AGN($z$)& \%($z=0$)  \\ 
\hline                    
   0.5-1.0 & $40 \,\, \pm 12 \,\,(31.4 \pm 9.4)$ & 30 (24) & 6 & 30 (24) & -- & -- & -- & -- \\  
   1.0-1.5$^\ast$ & $25.9 \pm 3.9 \,\,(24.8 \pm 3.8)$  & 35 (34) & 1 & 20 (19) & $8.1 \pm 2.2 \,(7.5 \pm 2.1)$ & 45 (42) & 3 & 40 (37) \\
   1.5-2.0$^\ast$ & $19.4 \pm 3.1 \,\, (17.0 \pm 2.9)$ & 45 (39) & 6 & 15 (13) & $7.3 \pm 1.9 \,(6.3 \pm 1.7)$ & 47 (41) & 6 & 36 (31) \\
   2.0-3.0$^\ast$ & $10.1  \pm 1.5 \,\,(\,\,\,8.9 \pm 1.4)$ & 53 (47) & 6 & 8 (7) & $3.0 \pm 0.8 \,(2.3 \pm 0.7)$ & 65 (50) & 15 & 15 (11)\\
\hline                  
\end{tabular}
\end{table*}
 
\begin{figure*}
\centering 
\includegraphics[width=15cm]{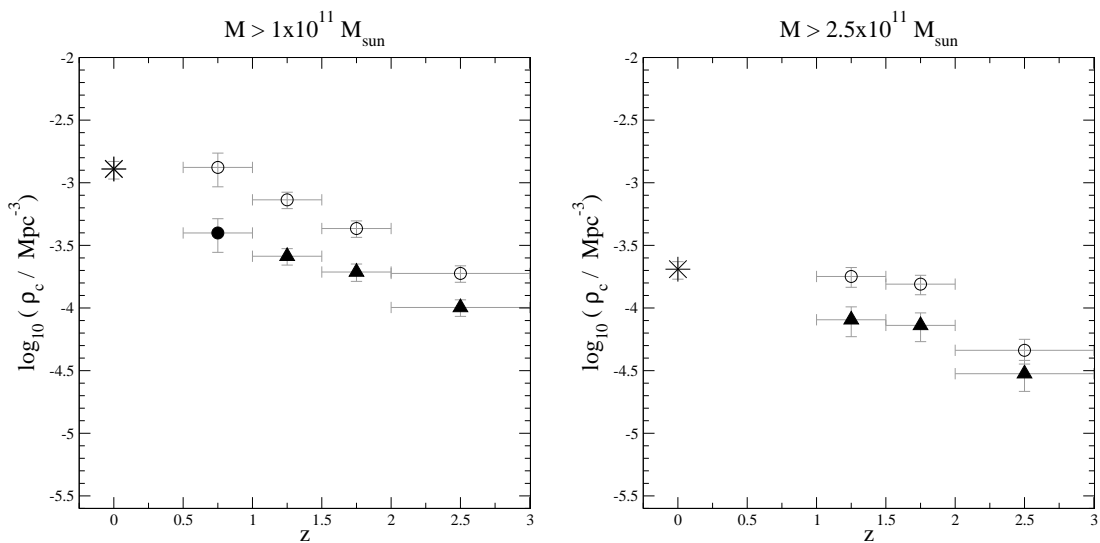}
\caption{The comoving number densities of galaxies with bolometric IR luminosity $L_{IR}>10^{11}\, L_\odot$  (filled symbols), with  assembled stellar mass  $M>1\times10^{11} \, M_{\odot}$ (left panel) and $M>2.5\times10^{11} \, M_{\odot}$ (right panel), versus redshift. Circles and upward-pointing triangles indicate directly derived  values and lower limits, respectively. Both star-forming galaxies and AGN/QSO are considered in all cases. The empty circles correspond to the densities of the total $K_s<21.5$ mag galaxy population with the same mass cuts (\cite{cap06a}). The asterisks at $z=0$ indicate the corresponding local values (Cole et al.~\cite{cole01}; Bell et al.~\cite{bell03}). 
             }
\label{cnumd11}
\end{figure*}


\section{The role of LIRGs and ULIRGs in stellar mass evolution}
\label{secrole}

    We now consider our  \tfm-detected $K_s<21.5$ mag sample to study how the number density of massive luminous IR galaxies evolved through cosmic time. Stellar masses have been obtained from the  optical/near-IR SED of each galaxy, which includes IRAC data at $3.6$ and $4.5 \, \rm \mu m$. The stellar masses are computed using the  rest-frame $K_s$-band luminosities, which are interpolated/extrapolated from the closest observed passband. We note that the use of IRAC data guarantees that we directly map the near-IR light of galaxies up to high ($z \sim 3$) redshifts. We estimate our stellar masses to be accurate within a factor $\sim 2$ (see ~\cite{cap06a} for more details).

    Bolometric IR luminosities ($L_{IR}$) for normal galaxies have been obtained using the Chary \& Elbaz~(\cite{chel01}) and Elbaz et al.~(\cite{elbaz02}) formulae, as in ~\cite{cap06b}. For the AGN/QSO present in our sample, we used the $L (12\,\rm \mu m)-L_{IR}$ relation derived by Takeuchi et al.~(\cite{tak05}), which makes no discrimination on the SED type. To interpolate/extrapolate the rest-frame $12 \, \rm \mu m$ flux of each AGN/QSO, we assumed that the SED of these objects followed a power-law $f_{\nu} \propto \nu^{\alpha}$ ($\alpha<0$). For each source, we derived the $\alpha$ value using its IRAC 3.6 and MIPS \tfm fluxes. 
    
     To assure that the  luminosities obtained with the Takeuchi et al.~(\cite{tak05}) relation were consistent with the Chary \& Elbaz and Elbaz et al. formulae,  we compared the obtained $L_{IR}$ values on our normal galaxies at $z \sim 1$, where the rest-frame  $12 \, \rm \mu m$ flux is directly mapped by the observed \tfm flux. We found that the luminosities obtained with the Takeuchi et al. relation are systematically lower by a factor $\sim 1.5$ with respect to the luminosities obtained with the  Chary \& Elbaz and Elbaz et al. formulae. Thus, in order to maintain the same luminosity scale for all of our objects, we multiplied the AGN/QSO luminosities derived with the Takeuchi et al. relation by a factor 1.5.

\begin{figure}
\centering 
\includegraphics[width=7cm]{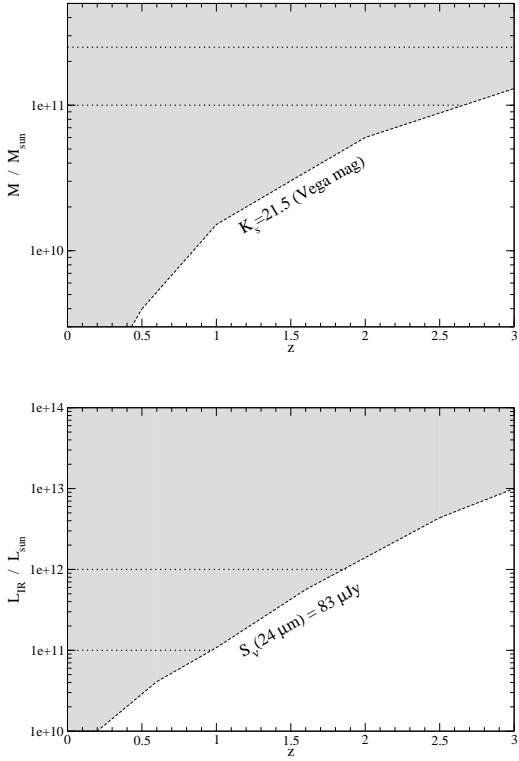}
\caption{ Upper panel: the evolution of the stellar mass completeness with redshift, as imposed by the $K_s=21.5$ mag cut. These completeness limits are strict, as they have been computed using a maximal mass-to-light ratio, corresponding to the template of a single stellar population formed at redshift $z \rightarrow \infty$. Lower panel: the evolution of the bolometric IR luminosity completeness with redshift, as imposed by the completeness limits of the \tfm survey. The luminosities in this plot have been computed using the Chary \& Elbaz and Elbaz et al. formulae, and extrapolated at $z<0.4$.  
}
\label{masslumcompl}
\end{figure}     
    
\begin{figure*}
\centering 
\includegraphics[width=15cm]{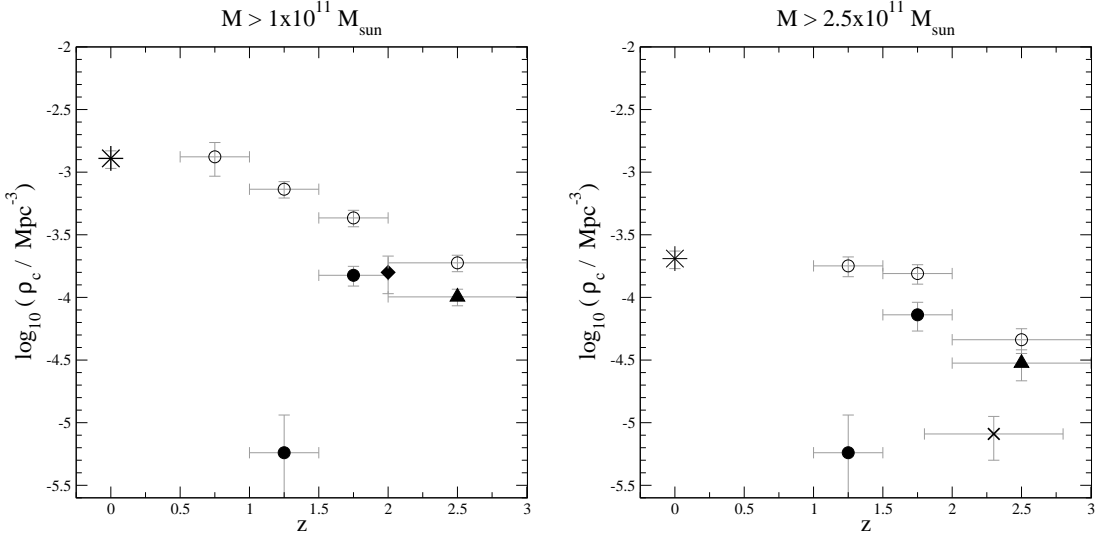}
\caption{The comoving number densities of galaxies with bolometric IR luminosity $L_{IR}>10^{12}\, L_\odot$.
 The references are the same as in Figure~\ref{cnumd11}. In the left panel, the diamond at $z=2$ shows the estimated density of ULIRGs obtained by Daddi et al.~(\cite{daddi05}). The cross-like symbol in the right panel indicates the number density of radio-detected sub-millimetre galaxies (Chapman et al.~\cite{chap03}).
             }
\label{cnumd12}
\end{figure*}

    Figure~\ref{cnumd11} shows the redshift evolution of the comoving number densities of galaxies with bolometric IR luminosity $L_{IR}>10^{11}\, L_\odot$ (filled symbols) with assembled stellar mass $M>1\times10^{11} \, M_{\odot}$ and $M>2.5\times10^{11} \, M_{\odot}$ (left and right panels, respectively). By definition, the IR luminosity cut adopted for these galaxies implies that all of them are either LIRGs or ULIRGs. Both star-forming galaxies and active nuclei have been considered for all the number densities  shown in Figure~\ref{cnumd11}. For star-forming galaxies, an IR luminosity $L_{IR}>10^{11}\, (10^{12})\, L_\odot$  implies a star-formation rate $SFR \gsim 17 \, (170) \, M_\odot \, \rm yr^{-1}$ (Kennicutt~\cite{kenn98}).  Circles in both panels of figure~\ref{cnumd11} correspond to directly derived values, i.e. number densities in the redshift bins in which we have completeness for all the $L_{IR}>10^{11}\, L_\odot$  galaxies (considering the  $S_\nu(24 \, \rm \mu m)=83 \, \mu Jy$ limit). Upward-pointing triangles indicate lower limits on the densities in the redshift bins where our sample is not complete for the adopted luminosity cut.  The evolution of the luminosity completeness imposed by the $S_\nu(24 \, \rm \mu m)=83 \, \mu Jy$ limit is shown in the lower panel of figure \ref{masslumcompl}.  For a comparison, the empty circles in both panels of Figure~\ref{cnumd11} show the densities of the total $K_s<21.5$ mag sample, as they were computed by \cite{cap06a}.  Analogous plots for only ULIRGs ($L_{IR}>10^{12}\, L_\odot$) are shown in Figure~\ref{cnumd12}.

    Our computed number densities  are $V_{max}$-corrected with respect to the $K_s=21.5$ mag magnitude limit, to encompass the computation of number densities for the total $K_s<21.5$ mag galaxy population. No corresponding $V_{max}$-corrections need to be applied to the \tfm band, as we considered all the \tfm identifications in our sample with no flux cut. Instead, we use the 80\% completeness limit of the \tfm catalogue ($S_\nu (24 \, \rm \mu m)= 83 \, \rm \mu Jy$) as a reference to determine up to which redshift the \tfm sample is complete for $L_{IR}>10^{11}\, L_\odot$ and $L_{IR}>10^{12}\, L_\odot$ galaxies.  We find that our \tfm sample is complete for LIRGs only up to redshift $z \sim 1$ and for ULIRGs up to redshift $z \sim 2$. Beyond those redshifts, we considered  our computed number densities to be lower limits on the real values.

  We note that mass completeness is not a concern in this analysis. The upper panel of figure \ref{masslumcompl} shows the evolution of the stellar  mass  completeness imposed by the $K_s=21.5$ mag cut. For stellar masses $M \gsim 1\times10^{11} \, M_{\odot}$, we are basically complete up to redshift $z=3$ (strictly $M > 1.3\times10^{11} \, M_{\odot}$ at $z=3$).

\begin{table*}
\caption{The comoving number densities of massive galaxies with bolometric IR luminosity  $L_{IR}>10^{12}\, L_\odot$ (ULIRGs) versus redshift. See caption of Table~\ref{tcnd11} for references.}  
\label{tcnd12}      
\centering          
\begin{tabular}{c r c r c r c r c}      
\hline\hline
 & \multicolumn{4}{c}{$\rm M>1.0\times10^{11} \,M_\odot$} & \multicolumn{4}{c}{$\rm M>2.5\times10^{11} \,M_\odot$} \\
\hline 
$z$  & $\rm \rho_c (\times 10^{-5}\, Mpc^{-3})$ & \%($z$) & \% AGN($z$)&\%($z=0$) & $\rm \rho_c (\times 10^{-5}\, Mpc^{-3})$ & \%($z$) & \% AGN($z$)& \%($z=0$)  \\ 
\hline                    
   0.5-1.0 & -- & -- & -- & -- & -- & -- & -- & -- \\  
   1.0-1.5 & $0.58 \pm 0.58 \,\,(0.58 \pm 0.58)$  & 0.8 (0.8) & 0 & 0.8 (0.8) & $0.58 \pm 0.58 \,\,(0.58 \pm 0.58)$ & 0.8 (0.8) & 0 & 0.8 (0.8) \\
   1.5-2.0 & $15.0 \pm 2.7 \,\, (13.1 \pm 2.5)$ & 35 (30) & 5 & 12 (10) & $7.3 \pm 1.9 \,(6.3 \pm 1.7)$ & 47 (41) & 6 & 36 (31) \\
   2.0-3.0$^\ast$ & $10.1  \pm 1.5 \,\,(\,\,\,8.9 \pm 1.4)$ & 53 (47) & 6 & 8 (7) & $3.0 \pm 0.8 \,(2.3 \pm 0.7)$ & 65 (50) & 15 & 15 (11)\\
\hline                  
\end{tabular}
\end{table*}

    The obtained number density values of $L_{IR}>10^{11}\, L_\odot$ and $L_{IR}>10^{12}\, L_\odot$ massive galaxies are listed in Tables~\ref{tcnd11} and ~\ref{tcnd12}, respectively. In the last redshift bin $z=2-3$, the depth of our \tfm catalogue only allows us to detect ULIRGs, so the lower limits we obtain for the densities of $L_{IR}>10^{11}\, L_\odot$ and $L_{IR}>10^{12}\, L_\odot$ galaxies are the same at these redshifts. The results presented in both tables show that luminous IR galaxies constitute a significant fraction of the assembled massive galaxies and this fraction increases with redshift. In particular, a minimum of $\sim 65\%$ of the most massive ($M>2.5 \times 10^{11}\, M_\odot$) galaxies already present at redshifts $z=2-3$ are ultra-luminous in the IR.

  Moreover, from comparison of figures~\ref{cnumd11} and ~\ref{cnumd12}, we can clearly confirm how the  role of IR galaxies among massive galaxies changes from being ULIRG-dominated at redshifts $z \gsim 1.5$ to be LIRG-dominated at $z \lsim 1.5$.   In effect, we observe that the number density of massive  ULIRGs  drastically falls below $z=1.5$. Within our sample, we find only one galaxy with $L_{IR}>10^{12}\, L_\odot$ and  $M>1\times10^{11} \, M_{\odot}$ at $1.0<z<1.5$, and none below these redshifts. ULIRGs might be present at lower redshifts, but are indeed very rare (e.g. Flores et al.~\cite{flo99}).
    
     Within our sample, the ratio of AGN/QSO among the ULIRGs with stellar mass $M>2.5 \times 10^{11}\, M_\odot$  reaches its maximum at $z=2-3$. This fact, as well as the high $SFR$ derived for the non AGN/QSO sources (\cite{cap06b}), highlight the importance of both star-formation and accretion activity in the evolution of the most massive galaxies already assembled at high redshifts. In contrast, the ratio of AGN/QSO  among the $L_{IR}>10^{11}\, L_\odot$ galaxies with  stellar mass $M>1.0 \times 10^{11}\, M_\odot$ reaches its maximum at lower redshifts $z=0.5-1.0$, but is a minor fraction (6\%) of all the $K_s<21.5$ mag galaxies with the same mass cut in the same redshift bin.

     With respect to the local population of galaxies with stellar mass $M>1.0 \times 10^{11} \, M_{\odot}$ ($M>2.5 \times 10^{11} \, M_{\odot}$), we find that a minimum of  20\%, 15\%, 8\%  (40\%, 36\%, 15\%)  had IR luminosities  $L_{IR}>10^{11}\, L_\odot$ by redshifts $z \sim 1.0, 1.5$ and $2.0$, respectively. 
     
     Galaxies with $L_{IR}>10^{11}\, L_\odot$ contain $\sim 24$\% of the total stellar mass density  at redshifts $z=0.5-1.0$ and a minimum of $\sim 35$ and $45$\%  at $z=1.0-1.5$ and $1.5-2.0$, respectively. The total assembled stellar mass densities are  80-90\%, 45-50\% and 25-30\%  of the local value at redshifts $z \approx 0.75, 1.25$ and $1.75$, respectively (Caputi et al.~\cite{cap05}; \cite{cap06a} and references therein). This implies that  $L_{IR}>10^{11}\, L_\odot$ galaxies contain $\sim$ 21\%, and a minimum of 16 and 11\% of the local stellar mass density at these respective redshifts.

     In the left panel of Figure~\ref{cnumd12}, we compare our computed number densities of ULIRGs with the estimate obtained by Daddi et al.~(\cite{daddi05}) from the analysis of $BzK$-selected galaxies (diamond-like symbol at $z=2$), and we find a very good agreement.
       
     Finally, in the right panel of Figure~\ref{cnumd12}, we show for reference the number density of radio-detected sub-millimetre galaxies with flux $S_\nu(850 \, \mu \rm m) \gsim 5 \rm mJy$ (Chapman et al.~\cite{chap03}; cross-like symbol in Figure~\ref{cnumd12}).  The number density of these galaxies appears to be smaller by a factor $\sim 4$ (2-3) than the density of  $M>2.5 \times 10^{11}\, M_\odot$  ULIRGs ($L > 5 \times 10^{12}\, L_\odot$ galaxies)  at similar redshifts. There is evidence that the majority of radio-detected sub-millimetre galaxies have luminous \tfm counterparts (Egami et al.~\cite{ega04}) and could be associated with the build up of the most massive galaxies (e.g. Stevens et al.~\cite{ste03}). However, the comparison of number densities shows that not all the most massive ULIRGs are associated with bright sub-millimetre galaxies (at least with  radio-detected ones). What kind of \tfm-selected ULIRGs are counterparts to sub-millimetre galaxies is still not completely clear (cf. Lutz et al.~\cite{lutz05}), and this is being investigated, for instance, as a part of the SCUBA Half Degree Extragalactic Survey (SHADES) project (Mortier et al.~\cite{mor05}).


\section{Linking the ULIRG and the ERG phases}

 \begin{figure*}
   \centering
   \includegraphics[width=16cm]{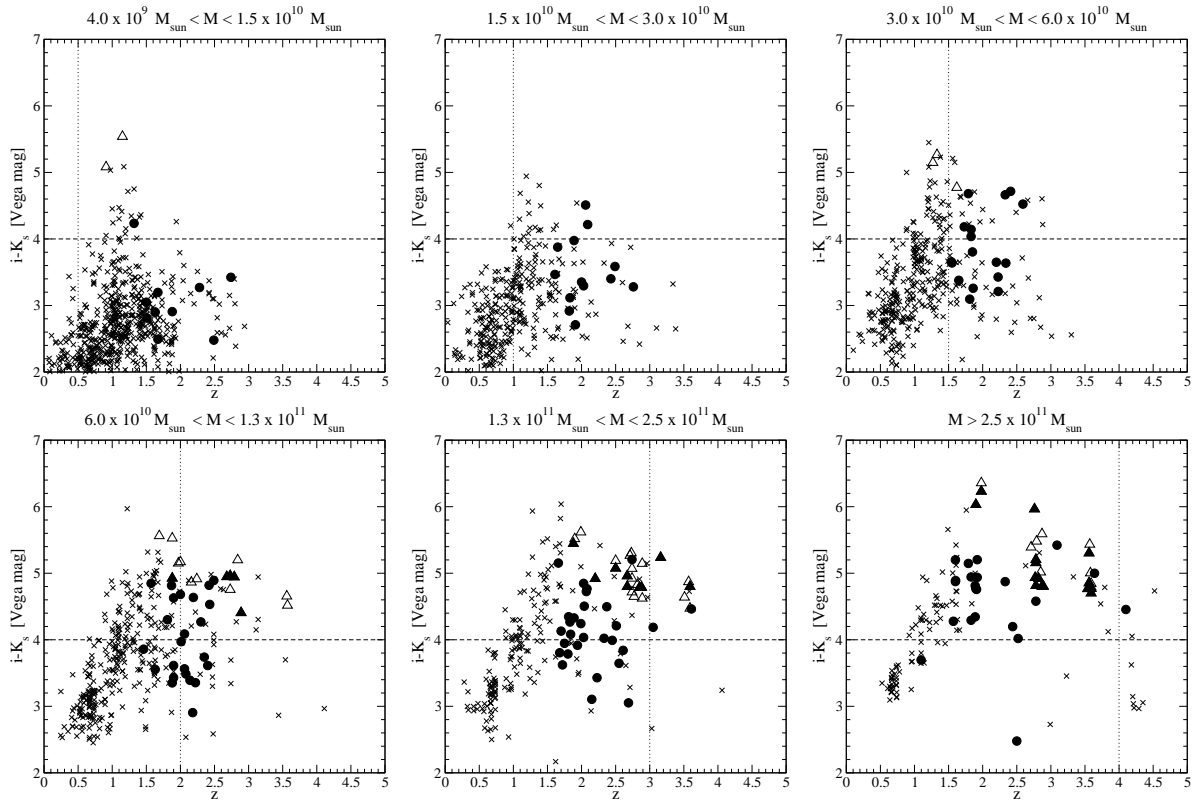}
      \caption{The $(i-K_s)$ colours versus redshift for the $K_s<21.5$ mag galaxies  in different stellar mass bins.   The $i$-band data corresponds to ACS/HST data. The filled symbols indicate \tfm-detected ULIRGs. All the triangles correspond to lower limits on the $(i-K_s)$ colours for sources below the $\sim 2\sigma$ detection limit in the $i$ band.   In each panel,  the dotted vertical  line shows the strict completeness redshift limit of the corresponding stellar mass bin. 
            }
      \label{imkcol}
   \end{figure*}

  The colours characterizing ERGs ($i-K_s>4$, Vega mag) are usually associated with the presence of old stellar populations  and/or starbursts heavily enshrouded by dust. \tfm data can be used to constrain the importance of these two components in the ERG population.

  Figure~\ref{imkcol} shows the $(i-K_s)$ colours of the $K_s<21.5$ mag galaxies of different stellar mass, with the \tfm-detected ULIRGs indicated by filled symbols.  Although formally the Chary \& Elbaz~(\cite{chel01}) and Elbaz et al.~(\cite{elbaz02}) relations do not allow us to compute bolometric IR luminosities beyond $z \sim 3$, the limits of the \tfm survey show that only ULIRGs can be detected beyond $z \sim 2$. Thus, every \tfm-detected source at $z>3$ is indicated as a ULIRG in Figure~\ref{imkcol}. Following our adopted classification criteria (cf. Section~\ref{secsep}), about one third of our \tfm-detected galaxies at $z>3$ are AGN/QSO. However, we note that our classification might be particularly incomplete at these high redshifts and the fraction of active galaxies among the \tfm-detected galaxies might be higher. Unfortunately, the present data do not allow us to better constrain the AGN fraction.

  We observe that ULIRGs are associated with  $K_s$ galaxies with a variety of stellar masses and $(i-K_s)$ colours at  redshifts $z \gsim 1.5$, with $\sim 60$\% of them having the characteristic colours of ERGs.

  Conversely, at $1.5<z<2$, $\sim$ 29\% of the $K_s<21.5$ mag ERGs are ULIRGs. At $2<z<3$, this percentage increases to a minimum of 49\%. If we consider only those ERGs with stellar mass $M>1.3 \times 10^{11} \, M_\odot$, these fractions are $40\%$ at $1.5<z<2$ and $\geq 52\%$ at $2<z<3$.  In particular, from Figure~\ref{imkcol}  we see that some of the reddest ERGs are not ultra-luminous in the IR (and some are not detected at all in our \tfm sample).  Put another way, no correlation seems to exist between the $(i-K_s)$ colours and the IR luminosity. This result appears to be independent of the assembled stellar mass. We note that the identification of ERGs with ULIRGs
is strictly complete up to redshift $z \approx 2$. Above that redshift, our \tfm sample starts to lose completeness even for ULIRGs, so there exists the possibility that some of the non-\tfm detected ERGs are also ultra-luminous in the IR.  The presence of old stellar populations might in part be responsible for the $(i-K_s)$ colour excess of those ERGs that are not ULIRGs. However, it is known that many of the reddest colours observed in ERGs can only be explained by the superposition of an evolved stellar component and non-negligible amounts of dust (Caputi et al.~\cite{cap04}; Papovich et al.~\cite{pap06}). 
  
  Daddi et al.~(\cite{daddi05}) find the average duty cycle for vigorous starburst with stellar mass $M \gsim 10^{11} \, M_\odot$ to be at least 50\% at redshifts $1.4<z<2.5$. In this context, the duty cycle is the fraction of time spent in the starburst phase over the total elapsed time in the redshift interval $1.4<z<2.5$.  The percentages of ULIRGs among massive ERGs we find here are consistent with  the duty cycle estimated by Daddi et al..  Recent studies  determined that  most of the brightest \tfm galaxies  at $z \gsim 2$  already have stellar masses $M \gsim 1 \times 10^{11} \, M_\odot$  and that star-formation activity  should mainly have proceeded by multiple discrete starbursts (\cite{cap06b}; Papovich et al.~\cite{pap06}). On the other hand,  the majority of the massive galaxies at  $z \gsim 1.5$ display extremely red colours. All these facts converge to a scenario in which most of the ERGs  --and not only the \tfm-detected ones-- could suffer a bright `mid-IR episode' (i.e. a ULIRG phase) up to several times in their lives. After a period of star-formation activity, a galaxy becomes fainter in the mid-IR, but it still would be an ERG.

  We compared the SEDs of our ERGs identified as ULIRGs, with those belonging to the other ERGs with similar stellar masses and  redshifts.  We found that the average properties (star formation history, age) of these SEDs are the same in the two groups of objects. Even the colour excesses are consistent within the error bars. The colour excesses are obtained from the convolution of the template SED with the Calzetti et al.~(\cite{calz00}) reddening law. For example, if we consider those ERGs with stellar mass $M>1.3 \times 10^{11} \, M_\odot$ at redshifts $2<z<3$ that are  ULIRGs and that  are not,  we find that the medians of the colour excess of the best-fit SEDs are $E(B-V)=0.30 \pm  0.07$ and $E(B-V)=0.22 \pm 0.05$, respectively. For the ERGs with the same mass cut but lying at redshifts $1<z<2$, the respective medians of the colour excess are $E(B-V)=0.17 \pm 0.05$ and $E(B-V)=0.10 \pm 0.07$. These results are basically the same if we divide the ERGs into those which have flux   $S_\nu(\rm 24 \, \mu m) > 83 \, \rm \mu Jy$ and $S_\nu(\rm 24 \, \mu m) < 83 \, \rm \mu Jy$ (including, of course, \tfm non-detections).
  
  Yan et al.~(\cite{yan04}) suggested that the \tfm-detected ERGs could be in the process of transforming into non-\tfm-detected early-type ERGs at $z \sim 1$. Our results are consistent with such a transformation hypothesis for the ERGs. More specifically: 1)  even if evolved stellar populations were already present at $2<z<3$, the $E(B-V)$ colour excesses characterising   massive ERGs  at these redshifts would be inconsistent with these objects being passive systems. It is possible that most of these objects are actually IR luminous, and some of them are non-detected within the limits of our survey. However,  if star formation indeed proceeds by discrete episodes, it will be more likely that a substantial fraction of the non-\tfm-detected ERGs are the detected ones in a post-starburst phase. 2) at $1<z<2$, the colour excesses are smaller on average, indicating that old stellar populations have a more important role in the colours of ERGs. However, some of the ERGs at these redshifts could still correspond to post-starbursts, as it has been demonstrated by spectroscopic studies (e.g. Doherty et al.~\cite{doh05}).


\section{Summary and discussion}
 
In this work we quantified the importance of luminous IR galaxies (LIRGs and ULIRGs) in the evolution of the massive $K_s$-band galaxy population. This allowed us to probe the role of star-forming galaxies (and quasar activity) in the history of stellar mass assembly.

Our results show that a substantial fraction of the massive galaxies ($M>1 \times 10^{11} \, M_\odot$) assembled at different redshifts are luminous in the IR. This fraction appears to be higher at high redshifts ($z \sim 2-3$) than at lower redshifts.  This result is clearly indicating that IR activity in massive galaxies --either due to star formation or AGN-driven--  has been much more important in the past.

We determined that a minimum of $53$ and $65$\% of the most massive ($M>1 \times 10^{11} \, M_\odot$ and $M>2.5 \times 10^{11} \, M_\odot$, respectively) galaxies present at  $z \sim 2-3$ were ULIRGs at those redshifts.

 Although the formation redshift of the first  most massive systems  cannot be determined from the depth of current surveys, our results show that the elapsed time between  $z=3$ and $z=2$ is of major importance in the construction of the bulk of the stellar mass in a significant fraction of massive galaxies. The number density of all the galaxies with stellar mass $M>1 \times 10^{11} \, M_\odot$ ($M>2.5 \times 10^{11} \, M_\odot$) rises by more than a factor two (three) between $z \sim 3$ and $z \sim 1.5$ (cf. fig.~\ref{cnumd11} and \ref{cnumd12}). This is entirely consistent with a major period of IR activity  at $z \sim 2-3$.

 The similarities between the optical/near-IR SEDs of massive ERGs that are ultra-luminous in the IR and those that are not  suggests that ULIRGs do not represent a particular type of near-IR selected galaxies. They rather could constitute a phase (the phase of high star-formation/quasar activity) during the life of a massive galaxy, and in particular during the period in which it is detected as an ERG.  Bright IR galaxies are only able to trace a fraction of the massive galaxies present in the Universe at a given time, but this fraction becomes very important ($\gsim 50$\%) at redshifts $z \gsim 2$.
 
%

\begin{acknowledgements}
    This paper is in part based on observations made with the \em{Spitzer} Observatory, which is operated by the Jet Propulsion Laboratory, California Institute of Technology, under NASA contract 1407. Also based on public GOODS datasets, obtained with the Advanced Camera for Surveys on board the Hubble Space Telescope operated by NASA/ESA and with the Infrared Spectrometer and Array Camera on the `Antu' Very Large Telescope operated by the European Southern Observatory in Cerro Paranal, Chile. We thank the GOODS teams for providing reduced data products.
    
    We thank George Rieke for reading the manuscript and the anonymous referee for making some useful remarks. KIC acknowledges CNES and CNRS funding. RJM acknowledges the support of the Royal Society. 
\end{acknowledgements}

\end{document}